\documentclass[a4paper,smallextended]{svjour3}       %
\usepackage{geometry}
\usepackage{url}
\usepackage{natbib}
\usepackage{amstext}

\usepackage{amsmath}

\AtBeginDocument{%
  }

\usepackage{tikz}
\usetikzlibrary{graphs,positioning,calc,decorations.pathreplacing,fit,shapes.multipart,shapes.geometric,snakes,scopes,backgrounds}
\usepackage{multirow}
\usepackage{listings}
\newcommand{\tool} {LeakGuard }
\newcommand{\toolend} {LeakGuard}
 \journalname{Empirical Software Engineering}

\begin{document}
\title{LeakGuard: Detecting Memory Leaks Accurately and Scalably} 
%
%

\author{
  Hongliang Liang \and
  Luming Yin \and
  Guohao Wu \and
  Yuxiang Li \and
  Qiuping Yi \and
  Lei Wang
}

\institute{
  Hongliang Liang (Corresponding author) \at
  TSIS Lab., Beijing University of Posts and Telecommunications \\
  \email{hliang@bupt.edu.cn}
  \and
  Luming Yin \at
  TSIS Lab., Beijing University of Posts and Telecommunications \\
  \email{lumingying@bupt.edu.cn}
  \and
  Guohao Wu \at
  TSIS Lab., Beijing University of Posts and Telecommunications \\
  \email{guohaowu@bupt.edu.cn}
  \and
  Yuxiang Li \at
  TSIS Lab., Beijing University of Posts and Telecommunications \\
  \email{liyuxiang@bupt.edu.cn}
  \and
  Qiuping Yi \at
  TSIS Lab., Beijing University of Posts and Telecommunications \\
  \email{yiqiuping@bupt.edu.cn}
  \and
  Lei Wang \at
  TSIS Lab., Beijing University of Posts and Telecommunications \\
  \email{wangcppclei@gmail.com}
}

\date{Received: date / Accepted: date}

\maketitle

\begin{abstract}
Memory leaks are prevalent in various real-world software projects,
thereby leading to serious attacks like denial-of-service.  
Though prior methods for detecting memory leaks made significant advance, 
they often suffer from low accuracy and weak scalability for testing large and complex programs. 
In this paper we present LeakGuard, a memory leak detection tool 
which provides satisfactory balance of accuracy and scalability.
For accuracy, LeakGuard analyzes the behaviors of library and developer-defined memory allocation and deallocation functions in a path-sensitive manner and generates function summaries for them in a bottom-up approach. Additionally, we develop a pointer escape analysis technique  to model the transfer of pointer ownership.
For scalability, LeakGuard examines each function of interest independently by using its function summary and under-constrained symbolic execution technique, which effectively mitigates path explosion problem. 
Our extensive evaluation on 18 real-world software projects and standard benchmark datasets demonstrates that LeakGuard achieves significant advancements in multiple aspects: it exhibits superior MAD function identification capability compared to Goshawk, outperforms five state-of-the-art methods in defect detection accuracy, and successfully identifies 129 previously undetected memory leak bugs, all of which have been independently verified and confirmed by the respective development teams.
\end{abstract}

\section{Introduction}

Memory leaks refer to situations where a memory
regions dynamically allocated by functions such as malloc, calloc, or the new
operator, is not properly released after their use. These unused memory
regions cannot be reclaimed by the operating system, leading to
depletion of system resources. Memory leak bugs in programs can be
maliciously exploited by attackers, potentially resulting in
denial-of-service attacks. Research shows that many small or uncommon memory leak defects can also lead to similar consequences~\citep{cantrill2003method}. As of July 2024, there are over 1,700 CVEs (Common
Vulnerabilities and Exposures) related to memory leaks. Memory leaks
have become a significant factor compromising the security of computer
systems.

Except for the standard functions like \texttt{malloc} or \texttt{free}, in real-world programs, developers often write their own MAD functions that manage multiple memory objects by invoking standard functions directly or indirectly. The identification of memory allocation/deallocation (MAD for short) functions is crucial in memory leak detection.
Prior studies make significant advances in detecting memory leaks though they are limited in either accuracy or scalability. For example, K-MELD~\citep{emamdoost2021detecting} assumes that developers often check the success of MAD functions, and has demonstrated its effectiveness on the Linux Kernel. 
SinkFinder~\citep{bian2020sinkfinder} identifies potential bugs or issues by focusing only on the data flow between specific pairs of functions.
MiROK~\citep{wang2023mining} mines abstract resource operation knowledge (Abs-RAR pairs) from Java code using rule-based and learning-based strategies, and then instantiates them into concrete RAR pairs to build a comprehensive pool for enhanced resource leak detection.
MLEE~\citep{wang2021mlee} assumes that  memory leak bugs often occur in early-exit (E-E) paths of a program.
It detects memory leak bugs by cross-checking the differences in
memory freeing operations between E-E paths and normal paths in Linux kernel.
Unfortunately, these assumptions are not always true in real-world software projects and hence limited their scalability. 

In contrast, NLP-EYE~\citep{wang2019nlp} employs data flow analysis and function summary
to accurately model customized MAD functions.
Based on NLP-EYE, Goshawk~\citep{lyu2022goshawk} can detect
Use-After-Free (UAF) and Double-Free (DF) vulnerabilities in large-scale software.
Goshawk generates summaries for MAD functions in a path-insensitive manner. Initially, it uses an AI model to filter functions related to memory allocation/deallocation based on function names and parameter lists. Then, it applies a path-insensitive data flow analysis method to identify memory management functions and generate summaries, including information about the memory objects allocated/deallocated by these functions. However, this approach may overlook some genuine MAD functions during the AI filtering stage. Moreover, due to the path-insensitive nature of the data flow analysis, it fails to effectively address pointer aliasing issues, leading to a higher rate of false positives in the generated function summaries. Moreover, it does not consider the path conditions of memory object allocation/deallocation, leading to a higher false positive rate in practical detection.
SMOKE~\citep{fan2019smoke} uses path-sensitive analysis to detect memory leak bugs, and then confirms the results via an SMT solver. However, its modeling of library functions is not precise, and it suffers from inaccurate pointer analysis, thereby resulting in over 40\%  false-negative ~\citep{aslanyan2022static}.

To mitigate the above issues, we present LeakGuard, a memory leak detection tool which provides satisfactory balance of accuracy and scalability. Specifically, we first use a path-sensitive approach to accurately identify MAD functions in the program and the memory objects managed by them, and then generate function summaries for these functions. Then we filter out the caller functions of the MAD functions for later symbolic execution. Finally we perform path-sensitive analysis on these caller functions via under-constrained symbolic execution ~\citep{ucse2015} to detect memory leak bugs.
Additionally, we develop a pointer escape analysis technique  to model the transfer of pointer ownership.
In extensive experiments conducted on 18 real-world software projects, including OpenSSL, MySQL, and the Linux Kernel, LeakGuard demonstrated superior capability in generating MAD function summaries compared to Goshawk and outperformed five state-of-the-art methods—Clang Static Analyzer (CSA), Infer, CppCheck, SMOKE, and SVF—in defect detection. Moreover, it identified 129 previously unknown memory leak bugs, all of which have been confirmed by the respective developers.

Our contributions are summarized as follows. 
\begin{itemize}
\item 
We first identify MAD functions in a bottom-up manner based on the function call graph and generate function summaries for each of them, thereby enhancing the accuracy. 
Next we analyze the caller functions of each MAD function by using under-constrained symbolic execution, which facilitates the scalability .
\item 
We propose a pointer escape analysis strategy for memory leak, which identify three
pointer escape scenarios as follows. A pointer pointing to a dynamic memory object is 1)  returned by a function;
2) assigned to function parameters; or 3) assigned to a global variable or a member of 
 a global structures. Accordingly we achieve effective results by employing data flow analysis technique.
\item 
We implemented the aforementioned techniques in LeakGuard and evaluated its effectiveness on 18 real-world software projects, including OpenSSL, MySQL, and the Linux Kernel. Experimental results demonstrate that LeakGuard surpasses the state-of-the-art method Goshawk in identifying MAD functions and outperforms five advanced memory leak detection tools—Clang Static Analyzer, Infer, CppCheck, SMOKE, and SVF—in defect detection. Furthermore, LeakGuard detected 129 previously unknown memory leak vulnerabilities, all of which have been confirmed by the respective developers.
\item 
We build a benchmark that contains 129 real memory leak bugs in real-world open-source software projects. We make our tool and benchmark publicly available to facilitate future researches.
\end{itemize}

\section{Motivation}

In this section, we describe the limitations of prior methods to detect memory leak by using a motivating example.

\begin{figure*}
\begin{minipage}{0.5\linewidth}
\begin{lstlisting}
// openssl/crypto/x509/v3_addr.c
static int make_addressPrefix(IPAddressOrRange **result, unsigned char *addr, const int prefixlen, const int afilen){
  ...
  IPAddressOrRange *aor = IPAddressOrRange_new();

  if (prefixlen < 0 || prefixlen > (afilen * 8))
    return 0;
  if (aor == NULL)
    return 0;
  aor->type = IPAddressOrRange_addressPrefix;
  ...
err:
  IPAddressOrRange_free(aor);
  return 0;
}

IPAddressOrRange *IPAddressOrRange_new(void){
  return (IPAddressOrRange *)ASN1_item_new(&IPAddressOrRange_it);
}
\end{lstlisting}
\end{minipage}
\hspace{1em}
\begin{minipage}{0.5\linewidth}
\begin{lstlisting}[firstnumber=last]
// openssl/crypto/asn1/tasn_new.c
ASN1_VALUE * ASN1_item_new(const ASN1_ITEM *it){
  ASN1_VALUE *ret = NULL;
  if (ASN1_item_ex_new(&ret, it) > 0)
    return ret;
  return NULL;
}

int ASN1_item_ex_new(ASN1_VALUE **pval, const ASN1_ITEM *it){
  return asn1_item_ex_new(pval, it);
}

static int asn1_item_ex_new(ASN1_VALUE **pval, const ASN1_ITEM *it) {
  ...
  switch (it->itype) {
    case ASN1_ITYPE_CHOICE:
      ...
      *pval = calloc(1, it->size);
      if (!*pval)   goto memerr;
  ...
}
\end{lstlisting}
\end{minipage}
  \caption{A memory leak in OpenSSL found by \toolend } 
  \label{fig:example-bug}
\end{figure*}

Figure \ref{fig:example-bug} shows a code snippet in OpenSSL. 
In the \texttt{make\_addressPrefix} function, a pointer
variable  \texttt{aor} points to a memory region returned by 
the function \texttt{IPAddressOrRange\_new} (Line 4). 
In fact, the functions \texttt{IPAddressOrRange\_new, ASN1\_item\_new, 
ASN1\_item\_ex\_new}, and \texttt{asn1\_item\_ex\_new} will be called in order before the
desired memory is allocated by the library function \texttt{calloc} (Line 37).
However, if the condition at Line 6 which has nothing to do with \texttt{aor} is true, 
the \texttt{make\_addressPrefix} function returns directly and thus results in a memory leak bug.



Unfortunately, three state-of-the-art static analysis tools, CSA, Infer, and CppCheck, failed to detect this bug. Firstly, these industrial-grade analysis tools prioritize efficiency. To address the path explosion problem and complete large-scale project analysis within time limits, they restrict path depth and node count during analysis and may use path-insensitive techniques. As a result, some deep-seated bugs, such as this memory leak, are overlooked.

Secondly, all three tools model memory allocation/deallocation (MAD) based on standard library functions. In the example code, the memory leak occurs in the \texttt{make\_addressPrefix} function, while the  \texttt{calloc}  library function is four levels away in the call chain. This call chain involves complex program semantics like loops and switches. The imprecise MAD behavior modeling in these tools thus decreases the analysis accuracy. 

Actually  this bug was found by \tool and confirmed by OpenSSL developers.

\section{Approach}
\subsection{Formalization of memory leak}
Below we formally describe the functions regarding memory and pointer
operations in Linear Temporal Logic (LTL), considering the ownership and
reference count of a memory region. 
The formula explanation of atomic propositions and operators is as follows.
\begin{itemize}
\item allocate(p): Allocates memory at address p.
\item free(p): Frees the memory pointed to by p.
\item owns(X, p): Entity X (e.g., caller, global variable) owns the memory at p.
\item transfer\_ownership(X, p): Transfers ownership of the memory at p to entity X.
\item shared(p, n): The reference count of the memory at p is n.
\item F($\phi$) (Eventually): $\phi$ will hold at some future time.
\item G($\phi$) (Globally): $\phi$ holds for all future time.
\end{itemize}

\subsubsection{Allocation Return with Ownership \& RefCount
(AROR)}\label{allocation-return-with-ownership-refcount-aror}

The function returns a dynamically allocated pointer, transferring
ownership to the caller and initializing the reference count to
\texttt{1}. The reference count is initialized to \texttt{1} on
allocation. Each copy of \texttt{p} increments the reference count
(\texttt{+1}), and each release decrements it (\texttt{-1}). Memory is
freed when the reference count reaches \texttt{0}. Its LTL formula is
defined as:\\
\begin{align}
\notag
 \exists p \cdot &\text{F}\left( \text{allocate}(p)
\land \text{shared}(p, 1) \land  \text{F}\left( \text{return}(p)
\land \text{transfer\_ownership}(\text{caller}, p) \right) \right) \\ \notag
&\land  \text{G}\left( \text{free}(p) \leftrightarrow \text{shared}(p, 0)
\right)
\end{align}

\subsubsection{Inner Allocation Return with Ownership \& RefCount
(IAROR)}\label{inner-allocation-return-with-ownership-refcount-iaror}

The function returns a pointer containing an inner structure allocation.
The parent pointer owns the memory, and reference counts manage cascade
deallocation. Deallocating the parent pointer \texttt{p} decrements the
reference count of the inner pointer \texttt{q} (\texttt{-1}). Memory
for \texttt{q} is freed when its reference count reaches \texttt{0}. Its
LTL formula is defined as:\\
\begin{align}
\notag
\exists p, q \cdot &\text{F}\left( \text{allocate}(q)
\land \text{shared}(q, 1) \land  \text{F}\left( \text{return}(p)
\land (q = p + \text{offset}) \land \text{owns}(p, q) \right) \right) \\ \notag
&\land  \text{G}\left( \text{free}(p) \rightarrow \left( \text{shared}(q,
n) \land \text{shared}(q, n-1) \right) \right)
\end{align}

\subsubsection{Deallocation of Argument with Ownership \& RefCount
(DAOR)}\label{deallocation-of-argument-with-ownership-refcount-daor}

The function takes ownership of an argument pointer and directly
releases its memory, resetting the reference count to \texttt{0}. The
function resets the reference count of \texttt{p} to \texttt{0} and
frees it. The caller relinquishes ownership of \texttt{p}. Its LTL
formula is defined as:\\
\[ \forall p \cdot \text{G}\left( \text{call}(f) \land \text{arg}(p)
\rightarrow  \text{F}\left( \text{shared}(p, 0) \land \text{free}(p)
\land \neg \text{owns}(\text{caller}, p) \right) \right) \]

\subsubsection{Deallocation of Argument Offset with Ownership \& RefCount
(DAOOR)}\label{deallocation-of-argument-offset-with-refcount-daoor}

The function releases memory at an offset of the argument pointer while
retaining ownership of the parent pointer. Only the offset pointer's
reference count is decremented (\texttt{-1}). The parent pointer
\texttt{p} retains its ownership and reference count. Its LTL formula is
defined as:\\

\begin{align}
\notag \forall p \cdot\ & \text{G}\left( 
  \text{call}(f) \land \text{arg}(p) \land \text{shared}(p + \text{offset}, n) 
  \rightarrow \right. \\
\notag & \left. \text{F}\left( 
  \text{free}(p + \text{offset}) \land 
  \text{shared}(p + \text{offset}, n-1) \land 
  \text{owns}(\text{caller}, p) \right) 
\right)
\end{align}

\subsubsection{Allocation in Global with Ownership \& RefCount
(AGOR)}\label{allocation-in-global-with-refcount-agor}

The function stores allocated memory in a global variable, managed via
reference counting. The global variable increments the reference count
(\texttt{+1}) when holding \texttt{p}. Memory is freed when the global
reference count reaches \texttt{0}. Its LTL formula is defined as:\\
\[ \exists p \cdot \text{F}\left( \text{allocate}(p)
\land \text{shared}(p, 1) \land  \text{F}\left(
\text{store}(\text{global}, p) \land \text{owns}(\text{global}, p)
\right) \right) \land  \text{G}\left( \text{free}(p)
\leftrightarrow \text{shared}(p, 0) \right) \]

\subsubsection{Deallocation of Global with Ownership \& RefCount
(DAGOR)}\label{deallocation-of-global-with-refcount-dagor}

The function releases memory held by a global variable and resets its
reference count to \texttt{0}. Its LTL formula is defined as:\\
\[ \forall p \cdot \text{G}\left( \text{owns}(\text{global}, p)
\rightarrow  \text{F}\left( \text{free}(p)  \land  \text{shared}(p, 0)
\right) \right) \]

\subsubsection{Argument Copy with  RefCount
(ACR)}\label{argument-copy-with-ownership-transfer-refcount-acaor}
The function copies a pointer to an argument or a global,  updating reference counts. 
The source pointer \texttt{p} increments its reference count (\texttt{+1}). 
Its LTL formula is defined as:
\[ \forall p, q \cdot \text{G}\left( \text{copy}(p, q) \rightarrow 
\text{F}\left( \text{shared}(p, n) \land \text{shared}(p, n+1) \right) \right) \]

By applying the above formulas,  our approach can 1) reduce false positives as ownership rules eliminate irrelevant
deallocation checks, 2) support for complex patterns because shared memory, nested structures, and
global ownership are handled rigorously.

\subsection{Overview}
\begin{figure*}
\centering

\centering
\includegraphics[scale=0.3]{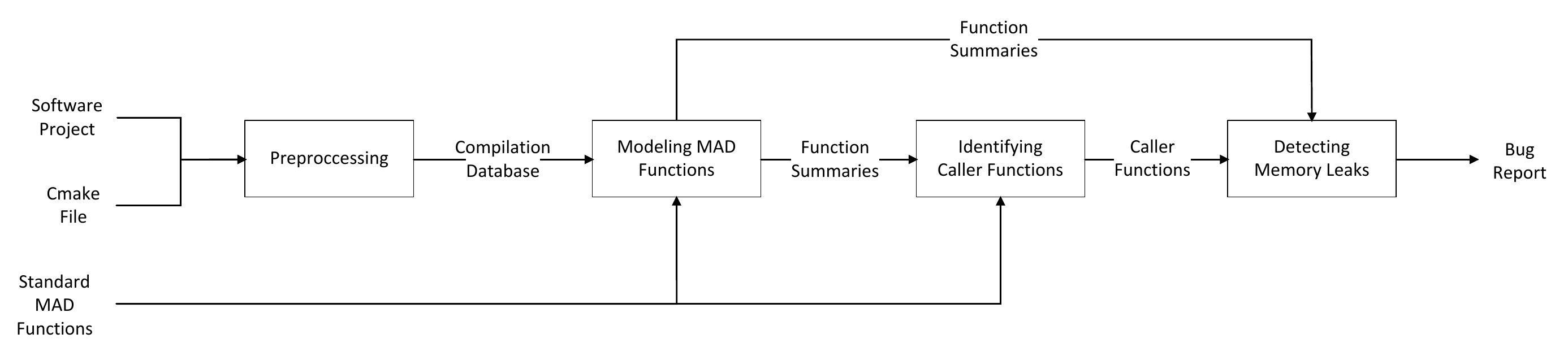} 
\vspace{-2ex}
\caption{Overview of LeakGuard} 
\label{fig:arch}
\end{figure*}
LeakGuard aims to detect memory leak in a scalable and accurate way.
Its workflow consists of four phases, as shown in Figure \ref{fig:arch}. 

Phase 1: Preprocessing the software project under test. 
Real-world software projects usually consist of
lots of code files and build files which contain rules and commands to build the whole project.
e.g., CMake/Make files. 
When building the project, LeakGuard records the compilation process 
into a compilation database for later use. 
thereby supporting for analyzing lots of open-source software projects.

Phase 2: Modeling MAD Functions. 
In real-world programs, an MAD function may be a standard library
function or a customized (i.e., developer-defined) one that allocates/deallocates memory. 
A library MAD function directly performs MAD operations, whereas a custom MAD
function may invoke a or multiple library MAD functions. 
Stand MAD functions are a prior and easily identified. 
We regard standard library MAD functions as predefined domain knowledge and encapsulate them within the tool as function summaries. To detect custom MAD functions, we implement an iterative process that starts by querying the call graph to identify all callers of known MAD functions. Subsequently, we employ path-sensitive data flow analysis to pinpoint MAD functions within these caller functions and refine their corresponding function summaries. This cycle continues iteratively until no additional summaries are incorporated.

Phase 3: Identifying Caller Functions of MAD functions. 
Based on the fact that a program doesn't expose memory leak if it has no 
MAD operations, the necessary condition for
a memory leak bug is a call at least to an MAD function. In other words, memory leak
bugs can only occur in functions that call an MAD functions. In
this step, we filter out all caller functions of the identified MAD functions by traversing the program call graph. 

Phase 4: Detecting Memory Leaks.
In this step, we perform path-sensitive analysis on these caller functions to detect memory leaks through under-constrained symbolic execution ~\citep{ucse2015}. Specifically, LeakGuard tracks the dynamic allocation and release of memory objects and uses a memory leak checker to detect related defects, and verifies the feasibility of bug-triggering paths via an SMT solver, thus ensuring detection accuracy. 
Moreover, we developed a pointer escape analysis mechanism based on data flow
analysis, which can effectively track the escape behavior of dynamic memory and
hence reduce false positives. In this way, LeakGuard effectively avoids the exploration of unnecessary 
paths and alleviate the path explosion problem, thereby ensuring its scalability.

\subsection{Modeling MAD Functions}\label{sec:model-mad}

Goshawk~\citep{lyu2022goshawk} abstracts the memory management behavior of MAD functions through function summaries and incorporates these summaries to model memory management behavior during symbolic execution, thereby effectively simplifying the path exploration process. Goshawk has successfully applied the aforementioned methodology to the detection of Use-After-Free (UAF) and Double-Free (DF) vulnerabilities, achieving promising results.
Compared to the aforementioned two types of defects, memory leak defects are more significantly impacted by inaccuracies in function summarization. For instance, when a function is mistakenly identified as a memory allocation function, this misclassification does not generate any false positives for UAF or DF defects. However, in the case of memory leak detection, each invocation of the erroneously classified function in the program will result in a corresponding false positive report of a memory leak. This heightened sensitivity to summarization inaccuracies underscores the critical need for precise function characterization in memory leak detection methodologies.
During the function summary generation phase, Goshawk first uses natural language processing to filter potential MAD functions based on function names and parameter information. It then employs a path-insensitive data flow analysis method to identify MAD functions within these functions. However, this method has certain limitations. Specifically, during the natural language processing phase, it may miss actual MAD functions. Additionally, the path-insensitive data flow analysis struggles to accurately model pointer assignment behaviors and has limited capabilities in pointer alias analysis. As a result, false positives and false negatives often occur when identifying memory objects managed by MAD functions.
Furthermore, in real-world application scenarios, most memory allocation and deallocation operations are closely related to runtime parameters, whereas Goshawk currently simplifies these memory management operations by assuming they are executed unconditionally. This oversimplification further reduces the accuracy of the analysis.

In light of the aforementioned considerations, the current function summarization methodologies prove inadequate for meeting the requirements of memory leak detection. Effective identification of memory leak defects necessitates precise function summaries that accurately capture and represent memory management behaviors.

We propose a bottom-up path-sensitive data flow analysis based on the function call graph to generate function summaries and obtain the complete set of memory objects managed by these functions.
The process of generating MAD function summaries is iterative. It begins with predefined MAD seed functions, which are stored in the system as prior knowledge. These seed functions are typically standard library functions used for memory allocation and deallocation. We traverse the call graph to identify the callers of these known MAD functions and perform path-sensitive data flow analysis on them. This analysis helps determine whether the callers qualify as MAD functions, and their function summaries are updated accordingly. The process continues iteratively until all MAD functions are identified.

In each iteration of the process, we determine whether the callers of the existing MAD functions are also MAD functions. To optimize analysis time and reduce redundant computations, we do not analyze all the callers of the already identified MAD functions. Instead, we focus on the callers of MAD functions that were newly identified in the previous iteration or those whose allocated or deallocated memory objects have increased. This targeted approach effectively streamlines the data flow analysis by concentrating on the most relevant functions.

In the process of generating memory allocation function summaries, we model the memory allocation behavior in the target function based on the currently identified memory allocation functions. At the same time, we track the propagation of dynamic memory throughout the program using path-sensitive symbolic execution techniques. When dynamic memory is returned as a function return value or passed as a function parameter out of the target function along any execution path, the current function is identified as a MAD function. We then record its name and the allocated memory objects and incorporate this information into the function summary.

In the process of generating memory deallocation function summaries, we adopt a path-sensitive approach to track each parameter within the target function. If, along any execution path, a parameter is passed to a known deallocation function recorded in the function summary, the target function is identified as a memory deallocation function. Subsequently, we record the name of the target function and the index of the memory object released within its parameters, and write this information into the function summary.

Our data flow analysis is both path-sensitive and field-sensitive, which means our tool can accurately distinguish memory allocation across different program paths while precisely identifying the nested relationships among members within structures.
Customized MAD functions and standard library MAD functions are not independent; instead, the memory management functionality of customized MAD functions is typically realized through the invocation of standard library MAD functions. In some cases, developers design customized MAD functions to maintain high cohesion by consolidating multiple memory allocation/deallocation operations with similar functionality into a single function. This design approach contributes to the widespread presence of customized MAD functions within the program.
Notably, our tool can effectively handle nested MAD functions, in other words, a 
memory object managed by an outer MAD function contains those managed
by all callee MAD functions.

Once a function is identified as an MAD function through data flow analysis, we store its name, characteristics, and the complete set of allocated/deallocated objects in the form of a function summary.
The function summary mainly include the function's name, its properties, and the set of objects it operates on. 
For instance, as shown in Figure \ref{fig:ffmpeg}, \texttt{dec\_alloc} is a custom MAD function in FFmpeg project, 
where a variable \texttt{dp} points to a memory region allocated by the function \texttt{av\_mallocz}. Subsequently,
its two members \texttt{dp->frame} and
\texttt{dp->pkt} point to memory regions allocated by the functions \texttt{av\_frame\_alloc} and
\texttt{av\_packet\_alloc}, respectively. Finally,  \texttt{dp} is assigned as the value of the function parameter \texttt{pdec}. 
In this way, the memory allocated by the \texttt{dec\_alloc} function can be passed to its caller functions.
Accordingly, the function summary generated
by LeakGuard is {\lstset{basicstyle=\normalsize\ttfamily}\lstinline|{"name": "dec_alloc", "function_type": "Allocator", "ret_objects": [], "para_objects": ["pdec", "pdec->frame", "pdec->pkt"]}|}. %

\begin{figure}
\centering
\begin{lstlisting}
// FFmpeg/fftools/ffmpeg_dec.c
static int dec_alloc(DecoderPriv **pdec, Scheduler *sch, int send_end_ts){
  DecoderPriv *dp;
  ...
  dp = av_mallocz(sizeof(*dp));
  if (!dp)  return AVERROR(ENOMEM);
  dp->frame = av_frame_alloc();
  if (!dp->frame)  goto fail;
  dp->pkt = av_packet_alloc();
  if (!dp->pkt)  goto fail;
  ...
  *pdec = dp;
   return 0;
fail:
  dec_free((Decoder**)&dp);
  return ret >= 0 ? AVERROR(ENOMEM) : ret;
}
\end{lstlisting}
\caption{Code snippet from FFmpeg project}
\label{fig:ffmpeg}
\end{figure}



\subsection{Identifying Caller Functions of MAD functions}

All MAD functions, including both standard and custom functions, have now been identified. Based on the fact that a function exhibiting memory leaks must call at least one allocation-type function, we can assume that a function that does not call any MAD functions will not exhibit memory leaks. In other words, we can safely exclude these functions. Therefore, at this stage, we first generate the call graph for the target program. Subsequently, based on the function call graph and MAD function summaries, we filter out the set of functions that do not need to be explored by the under-constrained symbolic execution engine.

The under-constrained symbolic execution technique allows us to freely choose entry points without starting exploration from the main function, which lays the foundation for excluding certain functions from exploration. Equally important as the under-constrained symbolic execution technique is the MAD function summarization technique. It abstracts memory allocation/deallocation behavior at a higher function level, helping us identify developer-defined MAD functions, which in turn allows us to exclude functions that do not perform memory allocation/deallocation, thereby reducing the exploration space for symbolic execution.

\subsection{Detecting Memory Leaks}

To check whether a caller function identified above contains memory leak,
\tool performs path-sensitive analysis on it via under-constraint symbolic execution ~\citep{ucse2015}.
Under-constrained symbolic execution allows exploration starting from specific functions, performing symbolic execution and constraint solving even in the absence of complete constraint collection. This approach effectively balances the accuracy and scalability of the detection.
Based on under-constrained symbolic execution, \tool directly executes a caller function of interest within the program, effectively skipping the costly path prefix from \texttt{main} to this function, which reduces the number and length of execution paths that must be explored. Additionally, LeakGuard leverages the ability of under-constrained symbolic execution to limit the exploration depth of function call chains, ensuring that only a manageable number of layers are analyzed for each function call. The combination of entry-point optimization and depth control significantly improves analysis efficiency while maintaining high coverage of code regions where memory leaks may occur.

Figure \ref{fig:ml-example-openssl} shows a code snippet in
OpenSSL, which contains a real memory leak bug. In the \texttt{frame\_ack} function, a pointer variable 
\texttt{ack\_ranges} points to a memory region allocated by
an MAD function \texttt{OPENSSL\_zalloc} (Line 8), which is deallocated at by the \texttt{OPENSSL\_free} (Line 15). 
However, \texttt{frame\_ack} function will exit if the condition at the line 12 is evaluated to be true, 
without releasing the memory region pointed to by \texttt{ack\_ranges}.
Based on the path-sensitive analysis, \tool detected this bug and reported the trigger condition: 
the \texttt{ossl\_quic\_wire\_peek\_frame\_ack\_}\\  
\texttt{num\_ranges} function returns zero and 
the \texttt{ossl\_quic\_wire\_peek\_frame\_ack\_num\_ranges} function also returns zero.
OPENSSL\_zalloc is a customed MAD function, which internally invokes standard library MAD functions to perform dynamic memory allocation. When LeakGuard encounters the OPENSSL\_zalloc function at the outermost layer, it can model the behavior of this function by reading the summary generated during the MAD summary generation phase, without the need to delve into the function itself. This method effectively reduces the complexity of data flow exploration and addresses the issue of standard library MAD functions being deeply embedded within the call chain, which prevents them from being explored and modeled.

\begin{figure}
\centering
\begin{lstlisting}
// openssl/ssl/quic/quic_trace.c
static int frame_ack(BIO *bio, PACKET *pkt) {
  OSSL_QUIC_FRAME_ACK ack;
  OSSL_QUIC_ACK_RANGE *ack_ranges = NULL;
  uint64_t total_ranges = 0;
  if (!ossl_quic_wire_peek_frame_ack_num_ranges(pkt, &total_ranges)
    || total_ranges > SIZE_MAX / sizeof(ack_ranges[0])
    || (ack_ranges = OPENSSL_zalloc(sizeof(ack_ranges[0]) * (size_t)total_ranges)) == NULL)
    return 0;
  ack.ack_ranges = ack_ranges;
  ack.num_ack_ranges = (size_t)total_ranges;
  if (!ossl_quic_wire_decode_frame_ack(pkt, 0, &ack, NULL))
    return 0;
  ...
  OPENSSL_free(ack_ranges);
  return 1;
}
\end{lstlisting}
\caption{Code snippet from OpenSSL project}
\label{fig:ml-example-openssl}
\end{figure}

Moreover, we establish a set of rules to identify memory leak bugs as follows:
 (1) When a pointer variable that points to a dynamically allocated memory
region goes out of scope and ceases to exist, the memory region it
points to has not been released. (2) The dynamically allocated memory
region pointed to by the pointer variable in Rule 1 has not been passed
outside the scope by being added to a global list, assigned to an
external parameter, or returned as a return value.

When a pointer variable exits its scope but the memory object it points to is not deallocated,  the first rule holds while this does not necessarily indicate a memory leak. This is because the memory object may have been reassigned to another variable before the pointer's scope ended. Specifically, ownership of the memory could have been transferred to a variable outside the scope prior to the pointer's destruction. According to the second rule, a memory leak is only considered if the allocated memory has not been reassigned outside the scope. This reassignment, known as pointer escape, can occur in three scenarios: (1) the memory is returned as a return value, (2) the memory is assigned to a function parameter or a global variable, or (3) the memory is added to a global data structure like a list. For the first two scenarios, we analyze the pointer's data flow to determine if it is returned or reassigned. In the third scenario, based on our observation that developers of real-world software often abstract such operations into independent functions with names including terms like add, insert or create and take as an argument the pointer to the global structure,  we analyze these functions independently to accurately model their behavior and identify potential pointer escapes.

The key to implementing the aforementioned rules is modeling MAD functions, as detailed in Section \ref{sec:model-mad}. LeakGuard models the behavior of MAD functions by reading the summaries generated during the MAD function modeling phase. In the MAD function summary generation phase, path-sensitive analysis is employed, which accurately models pointer assignment operations, thereby enhancing the accuracy of the generated summaries. However, the generated function summaries do not include conditional information related to memory management operations, as many of these operations depend on complex runtime parameters. Including such conditions in the summaries would significantly increase their complexity.

In real-world programs, many custom MAD functions do not perform memory allocation or deallocation unconditionally. Instead, these operations are often determined by runtime parameters. For instance, a MAD allocation function may allocate multiple memory objects, and developers can control which specific objects are allocated by passing allocation flags as arguments to the function. This approach is commonly employed when integrating multiple similar yet not entirely identical memory allocation requirements into a single MAD function, thereby enhancing code reusability and modularity.
During the memory leak detection process, the symbolic execution engine relies on function summaries to model memory management behaviors. Failure to incorporate path-sensitive information during the modeling of memory objects can lead to false positives in the detection results.

Based on the above analysis, it is crucial to account for the conditions under which memory management operations occur during the symbolic execution phase. If all memory management behaviors in an MAD function are unconditional, the symbolic execution engine will model these behaviors directly at the outermost layer based on the function summary. However, if some memory management operations in the summary are conditional, the engine will increase the exploration depth of the function call, collecting path constraints for each sub-MAD function invoked within the function. When the engine encounters an MAD function inside the function, it will recursively perform the above checks until path constraint collection is complete. Notably, the engine only performs deep exploration of the MAD function call chain. For other function calls within the MAD function that are not related to the invocation of sub-MAD functions, the engine does not increase the exploration depth. This feature is enabled by the under-constrained symbolic execution technique used by the engine, which allows flexible setting of exploration depth and enables the engine to utilize existing constraints to solve path feasibility when constraints are insufficient. In this way, the engine can collect the path constraints for each memory management operation without triggering path explosion issues, thereby accurately modeling memory management behaviors.

Specifically, during the MAD function modeling phase, we systematically record whether dynamic memory allocations or deallocations within the function are unconditional. When the symbolic execution engine encounters an MAD function where all memory objects are unconditionally allocated or deallocated, it directly models the corresponding memory management behaviors based on the function summary. However, if the MAD function summary indicates that certain memory management behaviors depend on runtime parameters, the engine will increase the exploration depth for both the function and its sub-MAD functions. It then performs constraint solving on the feasibility of memory management behaviors based on relevant path constraints, thereby eliminating infeasible memory management operations under the current path. This approach enables precise modeling of memory objects.

Previous methods, such as~\citep{shi2018pinpoint, fan2019smoke, emamdoost2021detecting}, typically stop tracking the data flow of a pointer once it is added to a global data structure and assume that such paths are not potential sources of memory leaks. Unlike these approaches, \tool offers a more precise modeling of the memory state of global variables, thereby improving the detection of memory leaks involving global data structures.

\section{Evaluation}
\subsection{Implementation}
LeakGuard is implemented based on LLVM 20.0.0 and Clang Static Analyzer (CSA) 20.0.0.
Specifically, we generate a function call graph by leveraging the RecursiveASTVisitor in the CSA  and generate function summaries by using its path-sensitive callback functions, such as checkPreStmt, checkBind, and checkPostCall,  in order to accurately capture a function's memory related behaviors. To identify functions where memory leaks may occur, we developed a filtering program in Python that reads the call graph and MAD function summaries, filters out the callers of MAD functions, and stores this information in text format in a local directory. Additionally, we incorporated a filtering mechanism into the CSA engine, which excludes functions unrelated to memory operations by reading and identifying the callers of MAD functions. Finally, a memory leak checker is developed  to read the generated function summaries,  perform pointer escape analysis, and enforce memory leak detection rules. 

\subsection{Research Questions}
In our empirical experiments, we want to answer the following research questions:
\begin{itemize}
\item RQ1: Is \tool effective in detecting memory leak bugs in the Juliet test suite and real-world software projects, compared to five state-of-the-art tools: CSA, Infer, CppCheck, SMOKE, and SVF?
\item RQ2:  Is LeakGuard efficient for testing large-scale complex software projects? 
\item RQ3: Can LeakGuard generate accurate function summaries compare to similar tools like Goshawk?
\end{itemize}

The experiments were conducted on an Intel Xeon server with 32 cores and 128
GB of memory, running Ubuntu 20.04. Each experiment was ran three times with 32 
 hours as time limit and average values are reported.

\subsection{Benchmarks and Baseline Tools}
As shown in Table \ref{tbl:found-leaks}, we use the latest versions (as of March 2025) of 18 widely used  software projects as our benchmarks, whose code base is over 70MLoC. 
On one hand, these programs are widely used and most of them have more than 10k stars on GitHub. 
On the other hand, they are extensively tested by other tools, e.g., Coverity~\citep{imtiaz2019developers} and SMOKE~\citep{fan2019smoke}.
Moreover, the types of  these programs contain process monitor, text editor,  browser, database and operating system kernel and so on, and their code sizes vary from 32 to 27,852 KLoC, which is a good reflection of software diversity.
Additionally, to accurately evaluate the detection capability of the testing tools, we selected the Juliet test suite as a benchmark. The Juliet C/C++ test suite is a systematically constructed collection of 64,099 small-scale test programs, covering 118 types of software weaknesses, including memory leaks, null pointer dereferences, unhandled exceptions, command injection, and deadlocks, among other common software defects. For our evaluation, we utilized the memory leak test cases from this test suite (i.e., the CWE-401 folder) as our test set.

We use CSA, Infer, and CppCheck as baseline tools to examine the 18 real-world programs mentioned above, as these open-source tools are capable of detecting memory leaks in various practical software projects and are actively under development.
To compare with SMOKE, we evaluated LeakGuard against the same benchmark programs used in SMOKE because SMOKE is developed based on an outdated version of LLVM (3.6), which cannot compile most of these programs successfully. 
Additionally, we selected CSA, Infer, and SVF as baseline tools to examine the test cases related to memory leak defects in the Juliet test suite. The effectiveness of these tools was evaluated by calculating their recall and false positive rates.

\subsection{RQ1: The effectiveness of LeakGuard}


We tested 18 real-world programs from the benchmark test set using CSA, Infer, CppCheck, and LeakGuard. Each detected bug was initially reviewed by two of the authors. Confirmed bugs were then reported to the corresponding developers through GitHub issues or other bug-report channels.
The numbers of confirmed bugs by developers are shown in Table \ref{tbl:found-leaks}. 
Note that 1) all bugs were discovered for the first time. 
2) for multiple memory leaks that are detected in different paths but point to 
the same memory region, we take them as the same bug and report only once.

\begin{table}[ht]
\centering
\caption{Memory Leak bugs Detected by LeakGuard, CSA, Infer, and CppCheck}
\label{tbl:found-leaks}
\begin{tabular}{lllrcccc}
\hline
Software & Type & Version & KLoC & LeakGuard & CSA & Infer & CppCheck \\ \hline
Htop & Utils & 3.3.0 & 32 & 1 & 1 & 0 & 0 \\ 
Tmux & Utils & 3.4 & 70 & 10 & 1 & 2 & 0 \\ 
Openssl & Network & 3.3.1 & 1226 & 26 & 6 & 3 & 0 \\ 
FFmpeg & Video/Audio & 7.0 & 1295 & 4 & 0 & 0 & 0 \\ 
MySQL & Database & 8.0.39 & 5640 & 20 & 8 & 5 & 0 \\ 
Blender & Graphics & 3.6.9 & 6321 & 7 & 2 & 0 & 0 \\ 
Wine & Emulator & 9.3 & 5597 & 6 & 4 & 0 & 1 \\ 
FireFox & Browser & 112.0.2 & 14349 & 3 & 3 & 2 & 0 \\ 
nanomq & Messaging & 0.21.10 & 519 & 8 & 2 & 5 & 0 \\ 
vim & Editor & 9.1.0099 & 414 & 13 & 0 & 0 & 0 \\ 
redis & Database & 7.2.4 & 204 & 2 & 0 & 0 & 0 \\ 
PHP & Interpreter & 8.2.19 & 1274 & 4 & 1 & 1 & 0 \\ 
swoole & Network & 5.1.3 & 138 & 1 & 1 & 1 & 0 \\ 
icu4c & Internationalization & 75 & 988 & 12 & 8 & 3 & 0 \\ 
zstd & Compression & 1.5.6 & 94 & 1 & 1 & 0 & 0 \\ 
FreeRDP & Network & 3.6.3 & 463 & 5 & 2 & 1 & 0 \\ 
godot & Game Engine & 4.4 & 4516 & 5 & 4 & 2 & 0 \\
Linux Kernel & Operating System & 6.9 & 27852 & 1 & 0 & 0 & 0 \\\hline
Total & - & - & 70992 & 129 & 44 & 25 & 1 \\ 
 \hline& 
\end{tabular}
\end{table}

Although these real-world software projects were tested extensively by well-known commercial tools like Coverity, LeakGuard detected 129 previously unknown memory leak bugs in these  projects, all of which were confirmed by the respective developers. In contrast, CSA, Infer, and CppCheck could only identify 44, 25, and 1 of the 129 bugs, respectively, further demonstrating the effectiveness and scalability of LeakGuard.

We analyzed the reasons why the three aforementioned comparison tools failed to detect the majority of memory leak defects. CppCheck uses a path-insensitive detection method in its analysis, which relies on syntax and program structure to analyze memory management behaviors. It does not differentiate between the specific conditions of each execution path in the program. As a result, CppCheck may not fully capture memory allocation/release behaviors under complex control flows. Additionally, CppCheck lacks a comprehensive cross-translation unit detection mechanism, whereas the data flows of the majority of the 129 defects involve multiple translation units.
CSA employs symbolic execution technology to detect memory leak defects. However, to balance detection efficiency, CSA typically imposes limitations on the depth of analysis during the detection process, particularly on the depth of function call chain analysis. Many of the aforementioned defects are triggered under relatively complex conditions, where the functions responsible for memory allocation and release operations involved in the memory leaks are deeply embedded within complex call chains. This structural complexity hinders CSA from accurately modeling the semantic behavior of these functions, thereby leading to the omission of such defects in the analysis results.
Infer employs abstract interpretation technology during program analysis. In its interprocedural detection process, Infer discards certain complex path constraints and cannot precisely handle the propagation of these constraints in complex contexts. Additionally, the tool approximates program statements during the detection process, leading to semantic loss. The errors introduced by such approximations increase sharply with deeper call chains and larger analysis scales, ultimately reducing the modeling precision.Therefore, when using Infer to detect the aforementioned defects, the tool may fail to identify those defects characterized by complex execution paths and triggering conditions spanning multiple functions due to semantic loss caused by program abstraction.

To facilitate future research, we have developed a memory leak detection benchmark that includes all 129 real bugs discovered by LeakGuard, as well as the false positives reported by the tool. These false positives have been cross-validated by three of the authors. Due to the application of under-constrained symbolic execution, \tool may generate a certain number of false positives caused by incomplete path constraint collection. However, the overall false positive rate remains below 40\%, and the defect reports provide detailed defect-triggering paths, ensuring a high level of user-friendliness.
For each real bug, the benchmark includes an HTML-formatted analysis report, which clearly presents the software version, the paths triggering the bug, feedback from the corresponding developers, and the patch to fix the bug.

SMOKE was built on LLVM 3.6, which cannot compile the latest versions of the 18  software projects in Table \ref{tbl:found-leaks}. As a result, we did not use SMOKE to analyze these programs. To compare the detection capabilities of LeakGuard and SMOKE, we applied LeakGuard to the test set used by SMOKE. The selection criteria for the test set are as follows:

\begin{itemize}
\item LeakGuard utilizes LLVM 20 as its compiler version. Since some test cases in SMOKE’s test set are incompatible with this version and cannot be compiled successfully, we excluded these cases from our evaluation.
\item   If a software project in SMOKE’s test set is already included in the 18 open-source programs we analyzed (albeit with a different version), we do not perform redundant testing. Instead, we directly use the defect detection results for its latest version. In this case, the number of defects detected by LeakGuard represents newly identified memory leaks in the software that were not detected by SMOKE, as all defects found by SMOKE in the tested version have already been fixed by the developers. There are two possible reasons why SMOKE failed to detect these defects: (1) the defects had not yet been introduced at the time of SMOKE's testing, or (2) SMOKE missed these defects. We have distinguished between these two cases in the test results.
\item   For the remaining test cases, LeakGuard analyzes the same software versions as those used by SMOKE. The detected defects are manually verified and compared with SMOKE’s results.
\end{itemize}

The defect detection performance of LeakGuard and SMOKE on the test set is presented in Table \ref{tbl:compare-with-smoke}.
The "Consistency" column indicates whether the software version used in LeakGuard’s analysis matches the version employed by SMOKE.
The "Exist" column quantifies the number of defects detected by LeakGuard that are present in the SMOKE test set. When the software version analyzed by LeakGuard aligns with that of the SMOKE test set, all defects identified by LeakGuard are inherently present in the SMOKE test set. In instances where the versions diverge, we meticulously examine the introduction timeline of each defect to ascertain its presence in the SMOKE test set or its subsequent introduction in later versions of the SMOKE test set.
When the tested version is consistent with SMOKE's version, since the software version is relatively outdated, the detected defects are manually confirmed by our team. When the versions are inconsistent, LeakGuard checks the latest version of the software, and the detected memory leak defects are reported to the developers for confirmation. When LeakGuard detects memory leak defects in real programs with versions inconsistent with the SMOKE test set, it indicates that the defect is newly discovered and was not detected by SMOKE, as all defects found by SMOKE have already been fixed in that version. 

The experimental results demonstrate that when analyzing software versions consistent with the SMOKE test set, LeakGuard not only achieves complete coverage of all defects identified by SMOKE but also detects additional memory leak defects.
A representative case is observed in httpd (version 2.4.29), where LeakGuard identified five memory leak defects, compared to only one detected by SMOKE.
This notable difference in detection outcomes clearly demonstrates LeakGuard's superior detection sensitivity and enhanced defect identification capabilities.
Moreover, for software versions inconsistent with the SMOKE test set, LeakGuard also demonstrated superior detection capabilities by successfully identifying multiple previously unknown defects.
A particularly compelling case is observed in OpenSSL, where SMOKE identified only four memory leak defects, whereas LeakGuard uncovered 26 previously undetected memory leak defects in the latest OpenSSL  (version 3.3.1) .
It is noteworthy that these 26 defects do not include the 4 defects detected by SMOKE, as these 4 defects had already been resolved in this version. Among these 26 defects, 5 were introduced during the SMOKE testing phase but were not successfully detected by SMOKE.
Through a systematic analysis of defect introduction timelines in version-mismatched scenarios, \tool  identified a total of 82 defects in the latest versions. Notably, 23 of these defects were present during SMOKE testing but remained undetected, while the remaining 57 defects were introduced in subsequent versions after the SMOKE test set. The successful detection of these defects by LeakGuard underscores its superior capability and advanced performance in defect identification.

To evaluate the false negative and false positive rates of LeakGuard, we selected test cases from the CWE-401 (memory leak) category of the Juliet Test Suite. This category comprises 1,364 independent test units, each consisting of one positive case and one negative case. For comparative analysis, we also included three widely used static analysis tools: Clang Static Analyzer (CSA), Infer, and SVF. The results of this evaluation are presented in Table~\ref{tbl:juliet-test-result}.

As shown in the table, LeakGuard achieved the highest performance among all tested tools, correctly identifying 1,336 out of 1,364 memory leak cases, with no false positives, and only 28 false negatives. In contrast, CSA, Infer, and SVF all exhibited significantly higher false positive and false negative rates. Specifically, CSA detected 490 true positives with 192 false positives; Infer detected 320 true positives and 250 false positives; and SVF found 520 true positives with 750 false positives.

LeakGuard produced no false positives during the evaluation, indicating its high precision and effective semantic modeling of MAD function behaviors. In contrast, the false positives observed in other tools are primarily due to incomplete or inaccurate semantic modeling of C/C++ features. For instance, Infer is unable to precisely evaluate path branch conditions during abstract interpretation, leading to semantic loss. CSA sacrifices precision in path condition evaluation for performance reasons, resulting in inaccurate judgments in complex branches or cross-translation-unit scenarios. Similarly, SVF employs sparse value flow analysis, which prioritizes memory-related semantics but lacks precision in other critical areas, leading to frequent misjudgments.

Regarding false negatives, we manually investigated the 28 cases missed by LeakGuard and found that these omissions were primarily due to limitations in the tool's analysis framework, which failed to cover rare or edge-case usage patterns. Notably, other tools also exhibited high false negative rates, with approximately 80\% of the false negatives attributed to unmodeled defect-triggering scenarios, particularly those involving complex interprocedural flows or edge-case resource handling.

The experimental results indicate that LeakGuard performs best on the Juliet Test Suite. Its high F1 score reflects a balanced performance in both precision and recall, achieving high recall while producing no false positives. LeakGuard outperforms the three comparative tools, all of which are constrained by incomplete or inaccurate semantic modeling of C/C++ features and path conditions.

\begin{table}[ht]
\centering
\caption{Testing Results of LeakGuard, CSA, Infer, and SVF on the Juliet Test Suite}
\label{tbl:juliet-test-result}
\begin{tabular}{lclcccc}
\hline
Tool & TP & FP & TN & FN & F1 Score \\ \hline
LeakGuard & 1336 & 0 & 1364 & 28 & 0.990 \\ 
CSA & 490 & 192 & 1172 & 874 & 0.479 \\ 
Infer & 320 & 250 & 1114 & 1044 & 0.331 \\ 
SVF & 520 & 750 & 614 & 844 & 0.395 \\ \hline
\end{tabular}
\end{table}

\begin{table}[ht]
\centering
\caption{Comparison of Detection Capabilities Between LeakGuard and SMOKE}
\label{tbl:compare-with-smoke}
\begin{tabular}{lclcccc}
\hline
Software & Type & Version & Consistency & LeakGuard & SMOKE & Exist \\ \hline
bftpd & FTP Server & 4.6 & Yes & 1 & 1 & 1 \\ 
htop & Utils & 3.3.0 & No & 1 & 1 & 1 \\ 
Memcached & Caching System & 1.5.4 & Yes & 1 & 1 & 1 \\ 
caffe & Machine Learning & 1.0 & Yes & 0 & 0 & 0 \\ 
LAME & Audio Encoding & 3.100 & Yes & 0 & 0 & 0 \\ 
zlib & Compression & 1.2.13 & Yes & 0 & 0 & 0 \\ 
tmux & Utils & 3.4 & No & 10 & 7 & 0 \\ 
httpd & Web Server & 2.4.29 & Yes & 5 & 1 & 5 \\ 
OpenSSL & Security & 3.3.1 & No & 26 & 4 & 5 \\ 
FFmpeg & Video/Audio & 7.0 & No & 4 & 0 & 2 \\ 
Godot & Game Engine & 4.4 & No & 5 & 6 & 0 \\ 
MySQL & Database & 8.0.39 & No & 20 & 14 & 8 \\ 
Blender & Graphics & 3.6.9 & No & 7 & 6 & 3 \\ 
Wine & Emulator & 9.3 & No & 6 & 20 & 4 \\ 
FireFox & Browser & 112.0.2 & No & 3 & 66 & 0 \\ \hline
Total & - &- &-&82&-&23 \\ \hline
\end{tabular}
\end{table}

\subsection{RQ2: The scalability of LeakGuard}

The scalability of LeakGuard essentially comes from its ability to
mitigate the path explosion problem during symbolic execution. The
stronger LeakGuard is at mitigating the path explosion issue, the
shorter the time required to detect the same project, thus enhancing
LeakGuard's scalability for large-scale programs. In our
work, under-constrained symbolic execution and the filtering of
potential memory leak functions based on the call graph are key to
achieving high scalability. Under-constrained symbolic execution helps
the tool avoid exploring predecessor paths, while filtering based on the
call graph allows symbolic execution to bypass exploring every single
function. 

In this experiment, we measured the time taken by LeakGuard and three baseline tools to detect the benchmark programs. 
Moreover, to demonstrate the contribution of filtering potential memory-leak functions, 
we disabled the module of identification of caller functions in LeakGurad (we call the variant as  LeakGuard-icf) and measured its time cost when analyzing all benchmark programs. 
Table \ref{tbl:compare-time} shows the experimental results.

\begin{table}[ht]
\centering
\caption{Comparison of Analysis time cost by LeakGuard, CSA, Infer, and CppCheck (seconds)}
\label{tbl:compare-time}
\begin{tabular}{lrrrrr}
\hline
Software & CSA & Infer & CppCheck & LeakGuard & LeakGuard-icf \\ \hline
Htop & 46 & 27 & 1 & 84 & 86 \\ 
Tmux & 274 & 128 & 27 & 209 & 333 \\ 
Openssl & 437 & 836 & 10 & 1117 & 1185 \\ 
FFmpeg & 651 & 1416 & 147 & 946 & 1507 \\ 
MySQL & 65316 & 18404 & 374 & 48204 & 75015 \\ 
Blender & 17036 & 6838 & 70 & 24664 & 33019 \\ 
Wine & 1695 & 3495 & 193 & 3530 & 5092 \\ 
FireFox & 22286 & 14957 & 4977 & 31811 & 37678 \\ 
nanomq & 64 & 157 & 2 & 141 & 146 \\ 
vim & 205 & 14498 & 109 & 320 & 419 \\ 
redis & 87 & 204 & 8 & 288 & 302 \\ 
PHP & 601 & 559 & 86 & 680 & 928 \\ 
swoole & 367 & 95 & 1 & 474 & 476 \\ 
icu4c & 749 & 353 & 31 & 876 & 916 \\ 
zstd & 162 & 75 & 4 & 118 & 191 \\ 
FreeRDP & 250 & 588 & 5 & 224 & 356 \\ 
godot & 51604 & 19762 & 498 & 6691 & 109970 \\ 
Linux Kernel & 28963 & 3069 & 252 & 14009 & 41874 \\  \hline
Total & 190793 & 85461 & 6795 & 134386 & 309493 \\
Average & 10600 & 4748 & 378 & 7466 & 17194 \\
Speedup & 1.42 & 0.64 & 0.05 & 1.00 & 2.30 \\
\hline
\end{tabular}
\end{table}

For each software project, LeakGuard finished its detection within 14 hours, ranging from 84 seconds
 (for Htop) at least to 13.4 hours (for MySQL) at most, which demonstrates the efficiency of our approach.
On average, LeakGuard is 1.42$\times$ faster than CSA, Infer is 1.56$\times$ faster than LeakGuard, and CppCheck is 20$\times$ faster than LeakGuard.
Specifically, CSA is faster than LeakGuard for 12 programs and slower for 6 programs, Infer is faster for 13 programs and slower for 5 programs, while CppCheck is faster than LeakGuard for all the tested programs.
The reasons behind are that 1) LeakGuard is slower than CSA  on the projects that have lots of caller functions of MAD functions, e.g., OpenSSL and Firefox, because LeakGuard spends more time on modeling the memory objects. 2) Infer is faster than LeakGuard on those large-scale projects, e.g., 
MySQL, Blender, Firefox and Linux Kernel, because Infer exploits abstract interpretation without symbolically executing the program like LeakGuard.
Infer's detection speed on the Wine project is similar to that of LeakGuard because not all files in Wine were analyzed. some files were ignored. In fact, the root cause is that Clang currently cannot built the entire Wine project.
3) CppCheck uses a path-insensitive detection method and limits its analysis to a single translation unit. This approach sacrifices detection capability in exchange for improved detection speed.
We observed that LeakGuard significantly benefited from the MAD function caller filtering feature when analyzing the Godot project, reducing the detection time by 94\% compared to LeakGuard-icf. Without considering the pruning effect of the MAD function caller filtering feature, LeakGuard-icf required relatively more time to analyze FFmpeg, Godot, MySQL, and Blender compared to other projects of similar scale. This is because these projects involve game engines, audio processing, and data processing, introducing a large number of graphics rendering, audio handling, and data modeling operations in the source code. The numerous macro definitions, interprocedural calls, and complex branching statements in the code further increased the detection time for the analysis engine.

Moreover, identifying caller functions contributes much to the efficiency of LeakGuard, resulting in a 0.43$\times$ (i.e., $1/2.30$) speedup. 
For example, \tool and \toolend-icf spent 3.9 hours and 11.6 hours on detecting Linux Kernel respectively. 
The reason behind is that only 36\% of all functions in Linux Kernel calls an MAD function.
Furthermore, \toolend-icf can also detect all of 129 real bugs found in the RQ1, which indicates that identifying caller functions is a satisfied optimization without hindering the detection effectiveness.


To evaluate the performance overhead of each stage in LeakGuard, we broke down and recorded the time consumption for four key analysis phases, as shown in Table \ref{tbl:compare-leakguard-stages}. These stages include Preprocessing, Modeling MAD Functions, Identifying Caller Functions, and Detecting Memory Leaks. The table presents the time spent on each stage across multiple software projects, along with the total time and percentage contribution of each phase.

As shown in the table, the memory leak detection stage dominates the total analysis time, accounting for 78.53\% of the overall runtime. This is because the stage involves symbolic execution and complex path exploration, which are time-consuming. The second most time-consuming phase is modeling MAD functions, taking up 18.26\% of the total time. This duration reflects the computational cost of summarizing memory-related behaviors within functions. In contrast, the time required for preprocessing and identifying caller functions is significantly lower, accounting for only 3.12\% and 0.10\% of the total time, respectively, indicating their limited impact on overall performance.

Notably, for software systems such as OpenSSL, PHP, and the Linux Kernel, a considerable portion of the total analysis time is spent on modeling MAD functions, primarily due to the large number of such functions present in these codebases. Overall, the module responsible for identifying the callers of MAD functions proves to be highly effective: it consumes only 0.1\% of the total analysis time while significantly reducing the detection time of LeakGuard by enabling early pruning of irrelevant paths. Nevertheless, even with the pruning optimization introduced by this module, the memory leak detection phase remains the primary performance bottleneck. Additionally, although the process of modeling MAD functions still incurs significant overhead, accounting for one-fifth of the total analysis time, this stage effectively enhances the tool's understanding of MAD function behavior, thereby improving the accuracy of detection.

\begin{table}[ht] 
\centering
\caption{Comparison of Analysis Time Cost by LeakGuard for Different Stages (in seconds)}
\label{tbl:compare-leakguard-stages}
\begin{tabular}{lrrrrr}
\hline
Software & Preproccessing & \parbox[t]{3cm}{Modeling MAD \\ Functions} & \parbox[t]{3cm}{Identifying \\ Caller Functions} & \parbox[t]{3cm}{Detecting \\ Memory Leaks} \\ \hline
Htop & 1 & 39 & 1 & 43 \\ 
Tmux & 1 & 74 & 1 & 133 \\ 
Openssl & 29 & 455 & 2 & 631 \\ 
FFmpeg & 13 & 427 & 3 & 503 \\ 
MySQL & 411 & 2511 & 10 & 45272 \\ 
Blender & 665 & 3422 & 13 & 20564 \\ 
Wine & 581 & 1932 & 19 & 998 \\ 
FireFox & 1075 & 4925 & 44 & 25767 \\ 
nanomq & 7 & 46 & 1 & 87 \\ 
vim & 12 & 93 & 1 & 214 \\ 
redis & 106 & 57 & 1 & 124 \\ 
PHP & 34 & 356 & 3 & 287 \\ 
Swoole & 17 & 99 & 1 & 357 \\ 
icu4c & 68 & 236 & 1 & 571 \\ 
zstd & 3 & 11 & 1 & 103 \\ 
FreeRDP & 12 & 96 & 2 & 114 \\ 
Godot & 682 & 3515 & 16 & 2478 \\ 
Linux Kernel & 471 & 6244 & 13 & 7281 \\ \hline
Total & 4188 & 24538 & 133 & 105527 \\ 
Percentage (\%) & 3.12\% & 18.26\% & 0.10\% & 78.53\% \\ \hline
\end{tabular}
\end{table}

\subsection{RQ3: The capability of LeakGuard in generating MAD function summaries}
The identification of MAD function summaries plays a critical role in the overall process by enabling more accurate recognition of a function's memory management behavior. This step helps optimize the analysis efficiency, allowing LeakGuard to focus on memory-relevant functions, thereby improving the scalability and precision of memory leak detection.

To evaluate LeakGuard's capability in MAD function summary identification, we compared it with Goshawk, a state-of-the-art tool to generate function summaries. The experiment was conducted on six real-world programs, and we compared the number of MAD functions identified by both tools.
The test results, as shown in Table \ref{tbl:custom-mad}, indicate that LeakGuard generates function summaries using a bottom-up approach based on the call graph and path-sensitive analysis. In the six real-world programs tested, LeakGuard identified 3,083 memory allocation functions and 3,631 memory deallocation functions, whereas Goshawk identified only 1,589 memory allocation functions and 1,469 memory deallocation functions.
LeakGuard identified 94\% more memory allocation functions and 147\% more memory deallocation functions than Goshawk. This improvement can be attributed to LeakGuard’s use of path-sensitive function summary generation, which effectively handles pointer aliasing issues, allowing for accurate tracking of pointer variables. In comparison, Goshawk’s function summary generation exhibits certain underreporting and false positives, leading to inaccurate results. Specifically, Goshawk fails to identify some true MAD functions during the NLP-assisted classification phase. Moreover, the path-insensitive traversal method employed in the data flow analysis phase fails to effectively handle complex program data flows and pointer aliasing issues, resulting in both underreporting and false positives, thereby reducing the accuracy of the results.

We identified two main reasons why LeakGuard was able to recognize more MAD functions than Goshawk: 1) Some MAD functions do not exhibit typical memory allocation/release characteristics in their names, which led Goshawk to exclude them from being classified as potential MAD functions for data flow analysis. 2) Goshawk's path-insensitive analysis method during data flow analysis is unable to track certain complex pointer data flows, preventing it from identifying these functions as MAD functions.

Figure 5 illustrates a memory allocation function named ACLMergeSelectorArguments in Redis, as identified by LeakGuard. In the fifth line of this function, a pointer variable named acl\_args is defined and allocated dynamic memory through the zmalloc function. Subsequently, the dynamically allocated memory region pointed to by this variable is returned at line 1131. Goshawk failed to identify this function due to its omission during the natural language processing phase, where it was not classified as a potential memory allocation function, thereby precluding any data flow analysis from being conducted on it.

Moreover, based on the identification of MAD functions, we improved \tool’s detection efficiency by excluding 408,550 (i.e., 32.7\%) function calls unrelated to memory operations, which do not invoke MAD functions. For instance, during the analysis of the Linux kernel, \tool excluded 64\% of the functions.

\begin{table}
\centering
\caption{Results of customized MAD functions and excluded functions during analysis}
\label{tbl:custom-mad}
\begin{tabular}{lrrrrrr}
\hline
\multirow{2}{*}{Program} & Allocators & Deallocators & Allocators &Deallocators\\ 
& (LeakGuard) & (LeakGuard) & (Goshawk) & (Goshawk)\\ \hline
Htop & 96 & 171 & 32 & 49 \\ 
Tmux & 406 & 362 & 38 & 95 \\ 
Openssl & 1445 & 1643 & 1058 & 970 \\
vim & 701 & 835 & 93 & 129 \\
zstd & 96 & 73 & 42 & 56 \\
redis & 339 & 547 & 326 & 170 
 \\  \hline
Total & 3083 & 3631 & 1589 & 1469
 \\  \hline
\end{tabular}

\end{table}

\begin{figure}
\centering
\begin{lstlisting}
// redis/src/acl.c
sds *ACLMergeSelectorArguments(sds *argv, int argc, int *merged_argc, int *invalid_idx) {
    *merged_argc = 0;
    int open_bracket_start = -1;
    sds *acl_args = (sds *) zmalloc(sizeof(sds) * argc);
    ...
    if (open_bracket_start != -1) {
        for (int i = 0; i < *merged_argc; i++) sdsfree(acl_args[i]);
        zfree(acl_args);
        sdsfree(selector);
        if (invalid_idx) *invalid_idx = open_bracket_start;
        return NULL;
    }
    return acl_args;
}
\end{lstlisting}
\caption{Code snippet from redis project}
\label{fig:ml-example-openssl}
\end{figure}

\subsection{Limitations}


Our approach may contain false positives due to the following reasons.
LeakGuard performs under-constrained symbolic execution independently on a function, without collecting path constraints  from the path prefix of the function. This approach reduces the exploration space and simplifies analysis while may miss relevant constraints from prior functions. From the viewpoint of detecting memory leaks, constraints from the path prefix are of limited benefit and significantly increase path complexity, though there are some cases where such constraints might be useful. Consequently, the insufficient constraints during symbolic execution may result in false positives.

Moreover, LeakGuard incorporates several optimization strategies to enhance its scalability for large-scale programs. For example, it limits loop unrolling to three iterations and avoids inlining functions that have over 100 basic blocks.  These strategies significantly reduce analysis time, whereas they may also lead to potential loss of program semantics, which may also result in false positives.


\section{RELATED WORK}

In this section, we discuss related work of detecting memory
leaks from two perspectives, i.e., static and dynamic detection methods.

\subsection{Static Memory Leak Detection}

Some tools, e.g., Sparrow ~\citep{jung2008practical}, Infer~\citep{harmim2019scalable}, employ abstract interpretation and model checking to detect multiples kinds of bugs like memory leak. 
Some methods leverage symbolic execution for memory leak detection. KLEE~\citep{cadar2008klee} is a well-known symbolic executor. By extending its MemoryObject module to record the allocation/release of dynamic memory objects, KLEE is capable of detecting memory leaks. 
Aslanyan et al.~\citep{aslanyan2024combining} first uses a context-sensitive, flow-sensitive, and field-sensitive static analysis to filter out potential memory leak bugs, then leverages KLEE ~\citep{cadar2008klee} to validate them.
SATURN~\citep{xie2005context} treats memory leaks as Boolean satisfiability problems and employs constraint solvers to detect memory leak bugs. 
Coverity~\citep{imtiaz2019developers} detects bugs in a program, including memory leaks, by utilizing data flow analysis, symbolic execution, and pattern matching. 
Several methods, e.g., SABER~\citep{sui2014detecting}, FASTCHECK~\citep{sui2012static}, PINPOINT~\citep{shi2018pinpoint}, and SMOKE~\citep{fan2019smoke}, simplify memory leak detection to the reachability problem on a sparse value flow graph. 
They focus on the data flow of heap objects and exclude the influence of irrelevant program statements. 
PCA~\citep{li2020pca} performs Andersen's pointer analysis ~\citep{andersen1994program} to derive points-to sets and constructs a call graph to recover indirect calls. To detect memory leaks, PCA builds a data dependence graph and checks if any reachable node from a memory allocation instruction has a corresponding "free" instruction; if not, it finds a memory leak.
Hector~\citep{saha2013hector} checks for memory leaks by identifying behavior inconsistencies within a function. It raises a warning when a resource is released on some paths of the function but not on others.
CLOUSEAU~\citep{heine2006static,heine2003practical} identifies memory leaks using a practical memory ownership model.

Abstract interpretation methods and graph reachability methods are based on a simplified program model for improving efficiency, which leads to inaccuracy as shown in our RQ1 experiment. Symbolic execution based techniques offer comprehensive exploration of programs and show better accuracy, though the challenge of path explosion limits their scalability as shown in our RQ2 experiment. By contrast, LeakGuard combines under-constrained symbolic execution and function summary to enhance its accuracy, and further leverages path pruning to avoid exploring irrelevant paths, thereby alleviating the path explosion problem.  



\subsection{Dynamic Memory Leak Detection}
Some methods ~\citep{hastings1992purify}~\citep{serebryany2012addresssanitizer}leverage source code instrumentation technique to detect memory leaks. For example,  Purify~\citep{hastings1992purify} inserts additional checking instructions directly into the source code of program under test, and tracks the usage of memory objects during execution. 
Some tools ~\citep{nethercote2007valgrind,bruening2011practical} exploits binary code instrumentation for memory leak detection. For instance, the memcheck tool in Valgrind ~\citep{nethercote2007valgrind}  analyzes a binary program via dynamic binary instrumentation and memory state tracking  during executing the program.
Dr.Memory~\citep{bruening2011practical} employs binary instrumentation to detect memory-related errors on both Windows and Linux platforms. 
Moreover, Sniper~\citep{jung2014automated} utilizes performance monitoring unit in CPU for instruction sampling and statistical analysis to track object staleness, enabling memory leak detection. 
Lee et al.~\citep{lee2014detecting} propose a deep learning framework that models heap objects and contains context information in order to detect memory leak bugs.

Compared to static analysis methods, it is difficult to achieve high coverage for dynamic methods. Additionally, generating inputs that trigger a potential bug is a challenge, especially when the triggering conditions are complex or even difficult to solve. 

\section{CONCLUSION}

We propose a memory leak detection method that balances 
accuracy and scalability well by combining  under-constrained symbolic execution, function summary and pointer escape analysis.
We implemented our method in a tool called LeakGuard and evaluated it on the latest versions of 18 well-known open-source software projects. Experimental results demonstrate that LeakGuard outperforms five state-of-the-art methods, namely CSA, Infer, CppCheck, SMOKE and SVF, while also surpassing Goshawk in terms of summary generation capabilities. LeakGuard completed the inspection of over 70 million lines of code within 37.3 hours and discovered 129 previously undetected memory leak bugs, all of which were confirmed by the corresponding developers.

\section*{Data Availability}
Our tool and benchmarks are publicly available at \url{https://gitee.com/yin-luming/leak_guard_tool}.


\section*{Conflict of interest}
The authors declare that they have no conflict of interest.

\bibliographystyle{spbasic}
\bibliography{reference}
\end{document}